\documentclass[11pt]{article}
\usepackage{moriond,epsfig}

\def\l {\lambda }

\def \ud {{1 \over 2} }

\def \bea {\begin{equation} }
\def \eea {\end{equation} }

\def \Eslash {E \kern-.5em\slash }
\def \pslash {p \kern-.5em\slash }
\def \kslash {k \kern-.5em\slash }

\def \mm {m_{3/2}}

\newcommand{\rpv}{\mbox{$\not \hspace{-0.10cm} R_p$ }}
\newcommand{\rpvi}{\mbox{$\not \hspace{-0.10cm} R_p$}}

\def\be{\begin{equation}}
\def\ee{\end{equation}}
\def\bea{\begin{eqnarray}}
\def\eea{\end{eqnarray}}

\begin{document}
\vspace*{4cm}
\title{THE COSMOLOGICAL GRAVITINO PROBLEM IN THE CONTEXT OF R-PARITY VIOLATION
\footnote{Invited talk given at the XXXVIIth {\it Rencontres de Moriond} 
session devoted to ELECTROWEAK INTERACTIONS AND UNIFIED THEORIES, 
March 9-16 2002, Les Arcs (France).}}

\author{ G. MOREAU }

\address{Physikalisches Institut, Bonn University,\\ 
Nussallee Strasse 12, D-53115 Bonn, Germany}

\maketitle\abstracts{Based on the R-parity violation option of the minimal
supersymmetric Standard Model, we
exa\-mine the scenario where the massive gravitino, relic from the hot
big-bang, is the lightest supersymmetric particle and can decay through
one or several of the trilinear R-parity violating interactions. We
calculate the rates of the gravitino decay via the various
three-body decay channels with final states involving three quarks
and/or leptons. By taking into account the present constraints on
the trilinear R-parity violating coupling constants and assuming
the gravitino and scalar superpartner masses do not exceed
$\sim 80 TeV$, it turns out that the gravitinos could easily have
decayed before the present epoch but not earlier than the big-bang
nucleosynthesis one. Therefore, the considered scenario
would upset the standard big-bang nucleosynthesis and we conclude that it
does not seem to constitute a natural solution for the cosmological
gravitino
problem.}

\section{Introduction}

In supergravity theories \cite{Nilles}, the gravitino, namely
the spin-$3/2$ supersymmetric partner of the graviton,
weakly interacts with all the particle species (including itself)
due to the small gravitational strength coupling $\sqrt{G_N}=1/M_P$,
$G_N$ being the gravitational constant and $M_P$ the Planck scale.
Hence, the gravitino-gravitino
annihilation rate is extremely small so that the gravitinos should
decouple at an early epoch of the universe history, and moreover
at an epoch characterized typically by a temperature $T_d$ higher than
the gravitino mass: $kT_d>m_{3/2}$ ($k$ being the {\it Boltzmann} constant)
\cite{Wein}. Therefore, the relic abundance of the gravitino should
be large, which is often denoted in the literature as the
cosmological ``gravitino problem''. One of the first solutions for the
gravitino problem to be envisaged is to compensate the
large gravitino relic abundance by a gravitino mass
sufficiently small, namely $m_{3/2}<1 keV$ \cite{Pagels}, to
respect the limit on the present universe energy density: $\Omega_0
\stackrel{<}{\sim} 1$.
For heavier gravitinos, a second type of available solution is by
shortening their lifetime so that they do not survive out at the late
epochs \cite{Wein}.
This can be realized in two characteristic options:
either the gravitino is not
the Lightest Supersymmetric Particle (LSP) and thus can decay into
an odd number of superpartners through gravitational and gauge
interactions (both of these ones couple an even number of superpartners),
or it is the LSP. A gravitino LSP can be realized within
various supersymmetric models, including the gauge mediated supersymmetry
breaking models \cite{Dine1,Dine2}, the models of low fundamental energy
scale \cite{Arkani} and even the conventional gravity mediated supersymmetry
breaking models (for some specific set of the supersymmetry breaking
parameters). If the gravitino is the LSP, it can decay
only into the ordinary particles of the Standard Model. Such a decay
channel must involve \cite{rev82} both gravitational and the so-called
R-parity
symmetry \cite{Salam,Fayet1} violating interactions (the latter ones couple
an odd
number of superpartners). The R-parity violating (\rpvi)
interactions are written in the following superpotential, in terms of the
left-handed superfields for the leptons ($L$), quarks ($Q$) and Higgs of
hypercharge $1/2$ ($H$) and the right-handed superfields for the charged
leptons ($E^c$), up and down type quarks ($U^c,D^c$),
\begin{eqnarray}
W_{\rpv}=\sum_{i,j,k} \bigg (\ud \l _{ijk} L_iL_j E^c_k+
\l^{\prime}_{ijk} L_i Q_j D^c_k+ \ud \l^{\prime \prime}_{ijk}
U_i^cD_j^cD_k^c
+ \mu_i H L_i \bigg ),
\label{superpot}
\end{eqnarray}
$i,j,k$ being flavor indices,
$\l_{ijk},\l^{\prime}_{ijk},\l^{\prime \prime}_{ijk}$
dimensionless coupling constants and $\mu_i$ dimension one parameters.
Note that in the scenario in which the gravitino is not the LSP,
the gravitino preferentially decays into an ordinary Standard Model particle
and its superpartner through
gravitational interactions, since the \rpv coupling constants
are severely constrained by the low-energy experimental bounds obtained at
colliders \cite{Drein,Bhatt}.

In the scenario of an unstable gravitino heavier than the LSP, the
gravitino decay can produce an unacceptable amount of LSP which conflicts
with
the observations of the present mass density of the universe \cite{Krauss}.
The scenario containing an unstable LSP gravitino, having decay channels
which involve \rpv coupling constants, is based on the violation of
the R-parity symmetry. Now, neither the grand unified theories,
the string theories nor the study of the discrete gauge symmetries give
a strong theoretical argument in favor of the conservation of the R-parity
symmetry in the supersymmetric extension of the Standard Model \cite{Drein}.
Hence, the scenario with an unstable LSP gravitino constitutes an attractive
possibility which must be considered as an original potential solution
with respect to the cosmological gravitino problem.

Our main purpose in the present work is to determine whether the
scenario of an un\-stable LSP gra\-vitino decaying via \rpv interactions
constitutes effectively a natural solution to the cosmological
gravitino problem. We will concentrate on the trilinear \rpv interactions,
namely $\l _{ijk} L_iL_j E^c_k$, $\l^{\prime}_{ijk} L_i Q_j D^c_k$ and
$\l^{\prime \prime}_{ijk} U_i^cD_j^cD_k^c$. With the goal of
enhancing the gra\-vitino instability, we will consider an optimistic type
of scenario in which several trilinear \rpv coupling constants have
simultaneously non-vanishing values. We will not consider the bilinear
\rpv term $\mu_i H L_i$ of Eq.(\ref{superpot}) which can be rotated
away by a suitable redefinition of superfields, for some specific choices 
of the \rpv soft supersymmetry breaking terms \cite{Drein,redef}.
The bilinear \rpv interactions as well as the pos\-sible
alternative of a spontaneous breaking of the R-parity symmetry have
been considered, within the context of the cosmological gravitino
problem, in a recent study \cite{Taka} which examines the two-body
gravitino decay mode into photon and neutrino. 
We also mention here the works \cite{Lee1}
and \cite{Lee2} in which have been considered the \rpv decay reactions
of single nucleon into a gravitino and a strange meson.

\section{Constraints in a scenario containing an unstable gravitino}

The equilibrium condition,
\begin{equation}
\Gamma \geq H(T),
\label{equi}
\end{equation}
$\Gamma$ being the gravitino decay rate and $H(T)$ the expansion rate of the 
universe, is reached at an epoch at which the cosmic energy density is dominated 
by the gravitino energy density $\rho_{3/2}$ \cite{Wein}.
Therefore, in determining the temperature $T_{3/2}$ at which the
equilibrium condition of Eq.(\ref{equi}) is reached (the gravitino decay
temperature) by solving the equation $\Gamma = H(T_{3/2})$, one must
take the following expression for the expansion rate,
\begin{equation}
H(T) \approx \sqrt{{8 \pi G_N \over 3} \rho_{3/2}},
\label{exp1}
\end{equation}
where $\rho_{3/2} = (3 \zeta(3) / \pi^2) (g(T) / g(T_d)) m_{3/2} (kT)^3$, 
$\zeta (x)$ being the {\it Riemann} zeta function ($\zeta(3)=1.20206\dots$)
and g(T) counting the total number of effectively massless degrees of freedom.
Hence, $T_{3/2}$ reads as,
\begin{equation}
kT_{3/2} \approx \bigg ({\pi \over 8 \zeta(3)}{g(T_d) \over g(T_{3/2})}\bigg
)^{1/3}
\bigg ({\Gamma^2 M_P^2 \over m_{3/2}}\bigg )^{1/3}.
\label{Tc}
\end{equation}
At the temperature $T_{3/2}$, most of the gravitinos decay. After the
gravitinos have decayed and their decay energy has been thermalized,
the temperature rises to the value $kT^\prime_{3/2}$ given by,
\begin{equation}
kT^\prime_{3/2} \approx {(90 \zeta(3))^{1/4} \over \pi}
\bigg ({m_{3/2} (kT_{3/2})^3 \over g(T_{3/2})}\bigg )^{1/4}
\approx  \bigg ( {45 \over 4 \pi^3} \bigg )^{1/4}
{g(T_d)^{1/4} \over g(T_{3/2})^{1/2}} \sqrt{\Gamma M_P}.
\label{Tcp2}
\end{equation}

In order to avoid large relic abundance of an unstable gravitino, this
one must decay before the present epoch. This requirement imposes
the following bound on the increased gravitino decay temperature
$T^\prime_{3/2}$,
\begin{equation}
T^\prime_{3/2} > 2.75 K \ \ (kT^\prime_{3/2} > 2.36 \ 10^{-10} MeV).
\label{bound1}
\end{equation}

Furthermore, the decay of the gravitinos causes an increase of the
temperature, which leads to an increase of the entropy density 
$s=(2 \pi^2 / 45) g(T) (kT)^3$
and hence to a decrease of the baryon-to-entropy ratio
$B=n_B/s$, $n_B$ being the baryon number density.
Therefore, if the gravitinos decay after the nucleosynthesis epoch, the
baryon-to-entropy ratio during the nucleosynthesis epoch must be much
greater than the present one, leading to the production
through nucleosynthesis of too much helium
and too little deuterium (compared with the constraints on the primordial
abundances derived from observational data) \cite{Wein}. In conclusion,
if the gravitinos decay before the present epoch, the temperature 
$T^\prime_{3/2}$, reached after the gravitino decay energy has been 
thermalized, has to be higher than the nucleosynthesis temperature, namely,
\begin{equation}
k T^\prime_{3/2} > 0.4 MeV.
\label{bound2}
\end{equation}

\section{Gravitino life time}
\label{sec:rate}

\begin{figure}
\psfig{figure=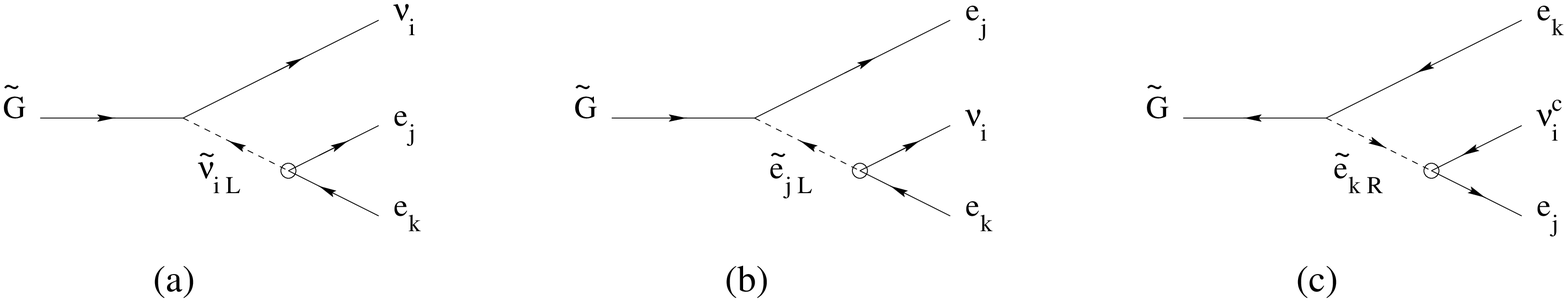,height=1.5in,width=15.cm}
\caption{{\it Feynman} diagrams of the gravitino three-body decay processes
involving the $\l_{ijk}L_iL_jE_k^c$ interactions. These interactions
are represented by the circled vertex. The other vertex correspond
to gravitational couplings. The arrows denote the flow of momentum of
the associated particles. Finally, $e$, $\tilde e$, $\nu$, $\tilde \nu$
and $\tilde G$ represent respectively the charged leptons, the charged
sleptons, the neutrinos, the sneutrinos and the gravitino, $i,j,k$
are generation indices, the $L, R$ indices stand for Left, Right
(chirality) and the $c$ exponent indicates a charge conjugated particle.
We have not drawn the charge conjugated reactions which can be
trivially deduced from these ones.
\label{grdec1}}
\end{figure}

The gravitino decay processes which involve the trilinear \rpv interactions
of Eq.(\ref{superpot}) and have the dominant rates are obviously the decays
through a virtual scalar superpartner into three ordinary fermions involving
one gravitational and one \rpv coupling. Those processes are represented in
Fig.\ref{grdec1} for the \rpv interactions of type $\l_{ijk}L_iL_jE_k^c$.
The reactions involving the $\l_{ijk}^\prime$ and $\l_{ijk}^{\prime \prime}$ 
couplings have similar structures.

As a function of the relevant
ratio of squared masses of the gravitino and exchanged sfermions,
$z = m_{3/2} ^2 /\tilde m ^2 $,
we find that the partial gravitino lifetime, associated to the decay
into a given flavor configuration of three light fermions
final state, decreases monotonically in the interval $0<z<1$ as,
\begin{equation}
\tau_{3/2} (\tilde G \to f_i f_j f_k)
\approx 10^9 - 10^{11} \ \bigg ( {1\over \hat \l_{ijk}^2} \bigg )
\bigg ( {1 TeV \over \mm} \bigg )^3  \ sec,
\label{lifetimeII}
\end{equation}
where $\hat \l$ stands for any one of the trilinear coupling
constants. For light gravitinos of mass
$m_{3/2} = {\cal O}(GeV)$ and $\hat \l_{ijk} = {\cal O}(1)$,
we see from Eq.(\ref{lifetimeII}) that
the gravitino lifetime exceeds by a few order of magnitudes
the age of the universe $t_0 \simeq 3.2 \ 10^{17} \ sec$.

\section{Numerical results and discussion}
\label{sec:disc}

\begin{figure}
\psfig{figure=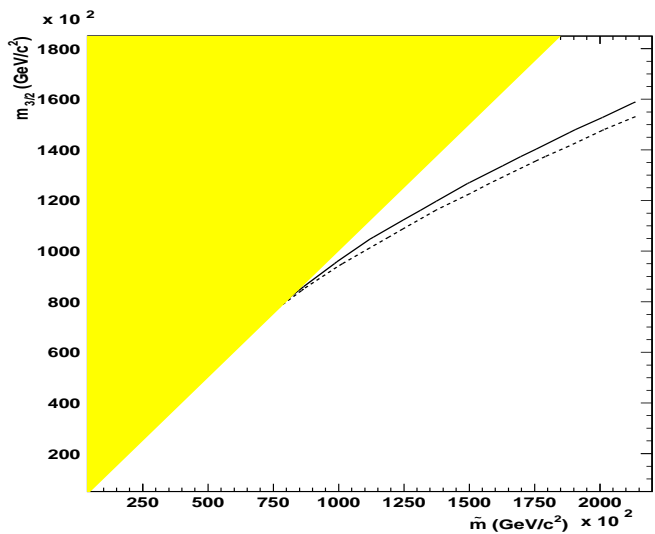,height=3.in,width=10.cm}
\caption{Domains of the $m_{3/2}(GeV/c^2)$-$\tilde m(GeV/c^2)$ plane
(gravitino versus superpartner mass) in which the gravitino decay
temperature is higher than the nucleosynthesis one, namely
$k T^\prime_{3/2} > 0.4MeV$. 
The region situated above the plain line corresponds to 
$k T^\prime_{3/2} > 0.4MeV$ in case the do\-minant \rpv coupling 
constant is $\l^{\prime \prime}_{213}=1.25$.
The domain situated above the dashed line corresponds to 
$k T^\prime_{3/2} > 0.4MeV$ in case the dominant 
\rpv coupling constants are $\l^{\prime}_{132}=1.04$, $\l^{\prime}_{211}=1$, 
$\l^{\prime}_{223}=1.12$, $\l^{\prime}_{311}=1.12$, 
$\l_{121}=1$ and $\l_{233}=1$. Finally, the colored 
region corresponds to the situation $m_{3/2}>\tilde m$ which must be considered
within a scenario where the gravitino is not the LSP.
\label{niv}}
\end{figure}

In Fig.\ref{niv} (plain line), we show the values of the gravitino and
superpartner masses for which $k T^\prime_{3/2} > 0.4MeV$. These values have
been
obtained by assuming that the single dominant \rpv coupling constant is
$\l^{\prime \prime}_{213}$ and is equal to its present limit \cite{Bhatt}.
\\ In Fig.\ref{niv}, we have set
all the scalar superpartner masses to a common value, denoted by
$\tilde m$. Moreover, the combination of statistical
factors $g(T_d)^{1/4}/g(T_{3/2})^{1/2}$, which enters Eq.(\ref{Tcp2}),
has been set to unity \footnote{Let us assume
tentatively that the thermalized degrees of freedom at the
gravitino decoupling and decay temperatures are the same as
those for the minimal supersymmetric Standard Model and
the present epochs, respectively. U\-sing then the
values $g(T_d) \simeq 915 / 4  \simeq 228.75$ and $g(T_{3/2}) =
43 / 11= 3.909$, one obtains: $T^\prime_{3/2} \propto
g(T_d)^{1/4}/g(T_{3/2})^{1/2} \simeq 1.96$. The actual numerical
value of this factor might be larger but not by very much.}.
Finally, the gravitino decay rate has
been multiplied by a factor of $2$ in order to count the charge
conjugated gravitino decay process and the masses of the final state
particles have been taken into account in the computation.
\\ We see in Fig.\ref{niv} (plain line) that as the superpartner mass
increases,
larger gravitino masses are needed to have $k T^\prime_{3/2} > 0.4MeV$. The
reason is that $T^\prime_{3/2}$ increases with the gravitino mass but is
suppressed if
the superpartner is getting heavier (since $T^\prime_{3/2}$ is proportional
to the squared root of the gravitino decay rate, as shown in Eq.(\ref{Tcp2})).

Let us consider now a type of scenario in which several
\rpv coupling constants, having some of the weakest present bounds,
are simultaneously non-vanishing and are equal to their present limit
obtained in the single dominant coupling hypothesis. 
Based on the strongest constraints on the products of
\rpv coupling constants and on the review \cite{Bhatt} of the present
limits on the single \rpv coupling constants, we find that the scenario
of this type leading to the highest gravitino decay temperature
$T^\prime_{3/2}$ (for $m_{3/2}=1TeV$ and $\tilde m=1.5 TeV$) corresponds
to the case where the simultaneously dominant \rpv coupling constants are,
\begin{eqnarray}
\l^{\prime}_{132},\ \l^{\prime}_{211},\ \l^{\prime}_{223},
\ \l^{\prime}_{311},\ \l_{121} \ \mbox{and} \ \l_{233},
\label{newd1}
\end{eqnarray}
which have the present bounds (obtained in the single dominant
coupling hypothesis)
$\l^{\prime}_{132}<0.34$ for $m_{\tilde q}=100GeV$ (and for instance
$\l^{\prime}_{132} \stackrel{<}{\sim} 1.2$ for $m_{\tilde q}=1TeV$)
\cite{Bhatt,Ellis},
$\l^{\prime}_{211}<0.06(m_{\tilde d_R}/100GeV)$ \cite{Barger,Led},
$\l^{\prime}_{223}<0.18(m_{\tilde b_R}/100GeV)$ \cite{Deba},
$\l^{\prime}_{311}<0.10(m_{\tilde d_R}/100GeV)$ \cite{Bhatt,Deba},
$\l_{121}<0.05(m_{\tilde e_R} /100GeV)$ \cite{Drein,Barger}
and $\l_{233}<0.06(m_{\tilde \tau_R}/100GeV)$ \cite{Drein,Barger}
\footnote{In Fig.\ref{niv}, the \rpv coupling constants of set \ref{newd1}
have been set to their perturbativity limit,
obtained from the requirement of perturbativity up to the gauge group 
unification scale \cite{aldedr}, since, in the whole intervale of $\tilde m$ 
covered by the figure, this latter limit is more severe than the corresponding 
present bound, which is a low-energy experimental constraint.}.
We note that no $\l^{\prime \prime}_{ijk}$
coupling can be added to the set \ref{newd1} of simultaneously
dominant \rpv couplings, since the experimental cons\-traints on
the proton decay rate force any product $\l^{\prime}_{ijk}
\l^{\prime \prime}_{i'j'k'}$ to be smaller than $10^{-9}$, in a
conservative way and for squark masses below $1 TeV$. This result
has been obtained in \cite{Smirnov} by calculating the proton
decay rate at one loop level. In contrast, we have checked
that no strong constraints exist on any product taken among the
\rpv couplings of the set \ref{newd1}.
\\ In this optimistic situation 
where several \rpv coupling constants are simultaneously present, 
the gravitino masses corresponding to $k T^\prime_{3/2} > 0.4MeV$ are smaller
than in the considered case of a single domi\-nant \rpv coupling constant 
(see Fig.\ref{niv}), since $T^\prime_{3/2}$ is typically proportional 
to the \rpv coupling constants and decreases with the gravitino mass.

The conclusion about Fig.\ref{niv} is that the gravitino mass, and thus the 
scalar superpartner masses, must exceed values of ${\cal O}(80 TeV)$, even
within the most optimistic scenarios, so that the condition of Eq.(\ref{bound2}) 
can be fulfilled. Nevertheless, this requirement conflicts with the constraint,
$\tilde m \stackrel{<}{\sim} {\cal O}(TeV)$, coming from the ``hierarchy problem'', 
namely the problem of natural coexistence of the electroweak symmetry breaking 
scale and the scale of new physics underlying the Standard Model (grand
unification scale, string scale \dots).

\section{Conclusion}

Along with the existence of the cosmic microwave background, big-bang
nucleosynthesis
is one of the most important predictions of the big-bang cosmology.
Furthermore, if
one assumes that the light nuclei (atomic number less than $7$) have
effectively been
produced through the big-bang nucleosynthesis, one finds that the
theoretical
predictions on the abundances of these light nuclei are in good agreements
with
the observational data \cite{Walker}. Now, as we have seen above, if one
believes that
the light elements have been synthesized through the standard big-bang
nucleosynthesis, the temperature $T^\prime_{3/2}$, reached after the 
thermalization of the gravitino decay energy, must be higher 
than the nucleosynthesis temperature. Therefore,
since we have found that this cannot happen in the scenario characterized
by an unstable LSP gravitino having a decay channel which involves trilinear
\rpv
interactions, this scenario does not seem to provide a realistic solution
to the
large relic abundance of the gravitino.

\section*{Acknowledgments}
The author is grateful to A.~Abada, P.~Fayet and J.-M.~Fr\`ere for 
invitation to the Moriond conference and interesting discussions.
This work was partially supported by the ``Alexander von Humboldt''
Foundation.

\end{document}